# Intertwined atomic-nanoscale-microscale structures via intralayer anisotropic Fe-chains in the layered ferromagnet FePd$_2$Te$_2$

Manyu Wang[1,2,+], Chang Li[1,2,+], Bingxian Shi[1,2,+], Shuo Mi[1,2], Xiaoxiao Pei[3], Shuming Meng[1,2], Yanyan Geng[1,2], Fei Pang[1,2], Rui Xu[1,2], Li Huang[3], Wei Ji[1,2], Hong-Jun Gao[3], Peng Cheng[1,2,*], Le Lei[4,*], Zhihai Cheng[1,2,*]

[1]*Key Laboratory of Quantum State Construction and Manipulation (Ministry of Education), School of Physics, Renmin University of China, Beijing 100872, China*

[2]*Beijing Key Laboratory of Optoelectronic Functional Materials & Micro-nano Devices, School of Physics, Renmin University of China, Beijing 100872, China*

[3]*Beijing National Laboratory for Condensed Matter Physics, Institute of Physics, Chinese Academy of Sciences, Beijing 100190, China*

[4]*Institute of Quantum Materials and Physics, Henan Academy of Sciences, Zhengzhou 450046, China*

**Abstract:** Controlling mesoscale and nanoscale material structures and properties through self-organized atomic behavior is essential for atomic-scale manufacturing. However, direct and visual studies on the cross-scale effects of such atomic self-organization on mesoscopic structures remain scarce. Here, we report the intertwined atomic-nanoscale-mesoscale structures via the intralayer Fe-chains in the sandwich-like layered FePd$_2$Te$_2$ crystal by scanning tunneling microscopy (STM) and atomic force microscopy (AFM). The hierarchical orthogonal corrugated morphologies are directly revealed and attributed to its chain-orientation-determined twinning-domain effect. Both Fe-chains of middle-sublayer and two kinds of Te atoms of top-sublayer are further atomically resolved at the sub-Å level, indicating the critical effects of Pd-atoms/voids on the intra-layer anisotropic Fe-chains and the interlayer structural alignment. The thermal-induced and strain-related structural transitions of surface layer are further investigated and discussed based on the proposed filling model of Pd-voids by the intralayer Pd-atoms. Our work not only provides deep understanding of this exotic layered magnetic material, and will inspire more perspectives for tailoring its anisotropic atomic-to-mesoscale structures and properties.

[+] These authors contributed equally to this work.
*Correspondence to Email: zhihaicheng@ruc.edu.cn  pcheng@ruc.edu.cn  lelei@hnas.ac.cn

**Introduction**

Two-dimensional (2D) materials have aroused extensive attention since the discovery of graphene [1]. Among them, anisotropic 2D material is one kind of 2D materials that possess different properties along different directions caused by anisotropic atoms' arrangement of the 2D materials, mainly including BP, borophene and low-symmetry TMDs [2-4]. This anisotropy may originate from the orientation of the lattice itself, such as in quasi-one-dimensional chain-like structures, or it may arise from more complex atomic structural domains or lattice distortions [5-8]. On this basis, they offer richer and more unique low-dimensional physics compared to isotropic 2D materials [9-13]. Among these, magnetic anisotropy is crucial for stabilizing long-range magnetic order in the two-dimensional limit [14]. Currently, most studied 2D magnets exhibit perpendicular magnetic anisotropy [15-23]. While larger perpendicular magnetic anisotropy allows for high-density information storage devices, two-dimensional magnets with easy-plane magnetic anisotropy are also highly attractive for various spin-related studies and applications. However, 2D magnets with in-plane anisotropy seem to be quite rare, besides $CrCl_3$ [24-26] and CrSBr [27-29].

Recently, a novel quasi-two-dimensional van der Waals material, $FePd_2Te_2$, has been discovered. Its one-dimensional Fe zigzag chains and strong in-plane easy-axis magnetic anisotropy make $FePd_2Te_2$ a unique material platform for studying both low-dimensional magnetism and spintronic applications [30]. Anisotropic magnetoresistance was performed to investigate the twin-induced in-plane anisotropy and four-fold symmetry was discovered. Hopkinson effect observed in both Ac and DC magnetic susceptibility indicates intense magnetic domain motion, which is rather rare in other van der Waals ferromagnets [31]. $FePd_2Te_2$ exhibits pronounced zero-field anomalous Hall effect (AHE) and anomalous Nernst effect (ANE) below $T_C$ (183K) making it a promising candidate material for practical thermoelectric spintronics applications [36]. It is noteworthy that studies reporting this compound indicate that the magnetic anisotropy of $FePd_2Te_2$ crystals is associated with significant twinning phenomena at the sub-micron scale [32-36], which is closely related to the one-dimensional Fe chains at the atomic scale. Consequently, $FePd_2Te_2$ provides a suitable platform for achieving controlled cross-scale interplay of structure and properties through atomic-level.

**In this work**, we systematically report and dissect the intertwined atomic-nanoscale-mesoscale structures that emerge from the intralayer anisotropic Fe-chains within the sandwich-like layered ferromagnet FePd$_2$Te$_2$. By combining scanning tunneling microscopy (STM) and atomic force microscopy (AFM), we directly visualize the hierarchical orthogonal corrugated morphologies from the mesoscale to the atomic scale. These unique morphologies arise from a self-organized twinning mechanism governed by Fe-chain orientation. At the atomic scale, our STM results achieve resolving the Fe-zigzag chains of the middle-sublayer and the two kinds of Te atoms of the top-sublayer at the sub-Å level. This indicating the critical effects of Pd-atoms/voids on the intra-layer anisotropic Fe-chains and the interlayer structural alignment. We further probe thermal- and strain-induced structural transitions of the surface layer, proposing a filling model of Pd-voids by intralayer Pd-atoms to explain the evolution. This work establishes a foundational framework for rationally designing and controlling its anisotropic structures and associated properties across atomic to mesoscale dimensions.

## Results

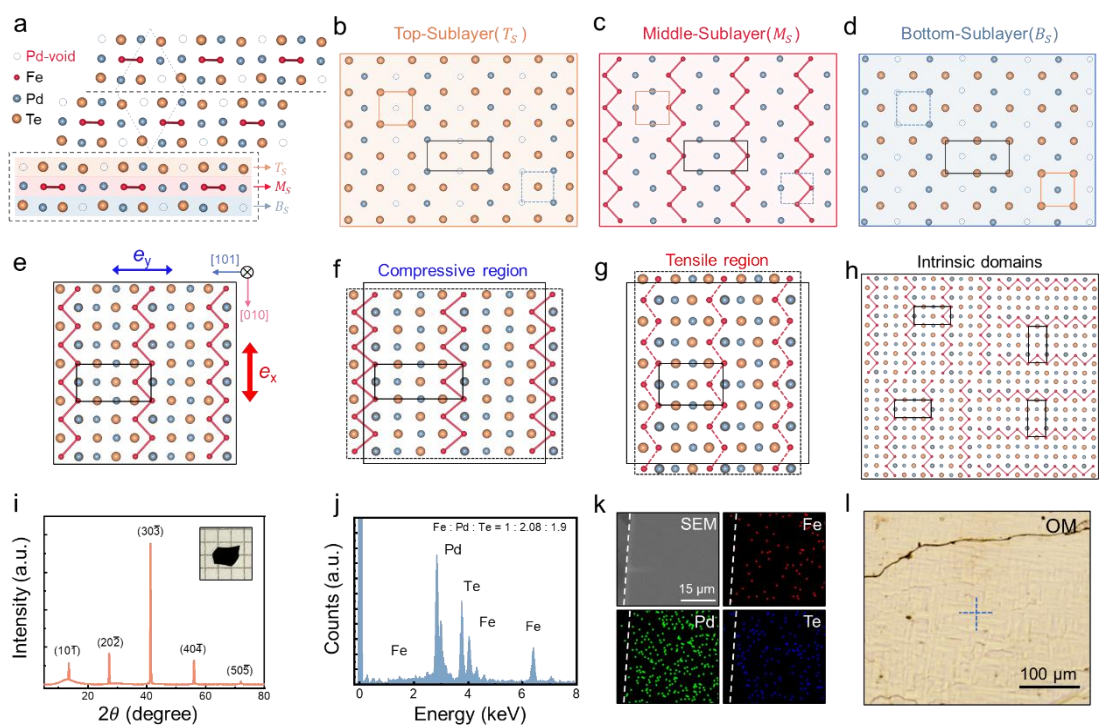

**Figure 1. Crystal structure of sandwich-like layered FePd$_2$Te$_2$.** (a) Side view of atomic structure of FePd$_2$Te$_2$. The crystal structure shows a layered structure along the [10$\bar{1}$] directions. The crystal can be easily exfoliated between the Pd$_{0.5}$Te sublayers, as indicated by the black dashed lines. (b-d) Top view of the top-sublayer (b, Pd$_{1/2}$Te), middle-sublayer (c, FePd), and bottom-sublayer (d, Pd$_{1/2}$Te) for the sandwich-like layer of FePd$_2$Te$_2$. (e) Top view of atomic structure of FePd$_2$Te$_2$. The Fe-zigzag chains lie in the exfoliation (10$\bar{1}$) plane and along the [010] directions, contributing anisotropic mechanical properties with large ($e_x$, along the chain) and small ($e_y$, vertical to the chain) Young's modulus. (f,g) Schematics of the compressive (f) and tensile (g) regions after structural relaxation from the unstressed layers of (e) due to their distributed anisotropic domains. The "compressive" and "tensile" refer to lattice distortions along the direction of the Fe-chains. The red dashed lines in (g) highlight regions where the Fe chains appear disrupted. (h) Schematic diagram of the coexistent two kinds of intrinsic domains with orientation-domain (O-domain) and phase-domain (P-domain). (i) XRD patterns from the cleavage plane of a FePd$_2$Te$_2$ single crystal. (j,k) EDS elemental analysis (j) and corresponding mapping (k) of one exfoliated flake. (l) Optical microscopy with orthogonal corrugated characteristics of FePd$_2$Te$_2$.

Figure 1a shows the anisotropic crystal structure of FePd$_2$Te$_2$, which is in monoclinic $P2_1/m$ symmetry with refined unit-cell parameters $a$=7.5024(5) Å, $b$=3.9534(2) Å, $c$=7.7366(7) Å, $\alpha = \gamma = 90°$, and $\beta = 118.15°$. As an atypical layered material, its cleavage interface (marked by dashed lines) occurs along the (10$\bar{1}$) plane between the sandwich-like layers. As shown in Fig. 1b-d, each sandwich-like layer of FePd$_2$Te$_2$ is made of top-sublayer ($T_s$, Pd$_{1/2}$Te), middle-sublayer ($M_s$, FePd) and bottom-sublayer ($B_s$, Pd$_{1/2}$Te), in which the inter-sublayer alignments of atoms are marked by the squares and rectangles. The anisotropic Fe-zigzag chains are confined by the Pd atoms within the middle-sublayer, and further sandwiched between the top- and bottom-sublayers. It is noted that the Pd$_{1/2}$Te-sublayers could be assumed as the PdTe-sublayers with the Pd-voids (marked by open circles) in the high-symmetric structure. The formation of Pd-voids in the PdTe-sublayer can be understood as the position-shifting of Pd atoms from the top- and bottom-sublayer to the middle-sublayer, which play important roles in the formation of Fe-chains and the relatively strong interface bonding, the proposed schematic models as shown in Figure S1.

The in-plane unit-cell of FePd$_2$Te$_2$ layers is marked by the solid rectangle in their top-view structural model of Fig. 1e, demonstrating the Fe-chain-induced anisotropic structures. The atomic bonding is expected to be stronger along the chains ($e_x$, parallel to the [010] direction) and weaker between the chains, leasing to the relatively large and small Young's modulus respectively, as illustrated in Figure 1e. Considering their anisotropic Young's modulus, the formation of compressive (C) and tensile (T) regions could occur under the internal strains after structural relaxations of spatially distributed domains (Fig. 1h and Fig. S2), as shown in Figure 1f and Figure 1g. The twinning-like orientation-domains (O-domains) originate from the low-symmetric anisotropic Fe-chains (middle-sublayer) and the quasi-high-symmetric PdTe (top-/bottom-sublayer), as shown in Figure 1h. While another kind of CDW-like phase-domains (P-domain) derive from the large quasi-supercell of Fe-chains with respect to the small quasi-cell of PdTe-sublayer, highlighted by the marked rectangles in Figure 1h. It is noted that these intrinsic domains are structurally equivalent and energetically degenerate, and should coexist within the synthesized bulk crystals. Both O-domains and P-domains exhibit domain boundaries at different angles, as illustrated with the structural model in Figure S3 and S4, which contribute complexed structure, morphology and properties.

The single-crystal samples of $FePd_2Te_2$ were grown by melting stoichiometric elements and characterized by X-Ray Diffraction (XRD) and Energy Dispersive Spectroscopy (EDS). By performing XRD measurements on the cleavage plane of $(10\bar{1})$, the Bragg peaks are indexed as $(H0\bar{H})$, as shown in Figure 1i, with an inset photo of typical crystal. The further EDS elemental analysis (Fig. 1j) and mapping characterizations (Fig. 1k) confirm its chemical composition of $FePd_2Te_2$ with a uniform spatial distribution. The optical microscope image of one typical crystal sample is given in Figure 1l, which clearly displays impressive orthogonal characteristics with tens of micrometer scale, clearly different from the known isotropic or anisotropic 2D materials. Notley, these distinct mesoscale features are highly reproducible across multiple samples and cleaved flakes, and should intrinsically origin from its sandwich-like layered atomic structures via the anisotropic Fe-chains between the quasi-symmetric PdTe-sublayers. To further explore the underlying mechanisms, the atomic force microscopy (AFM) and scanning tunneling microscopy (STM) measurements are performed to characterize the hierarchical structures across the atomic-nanoscale-mesoscale.

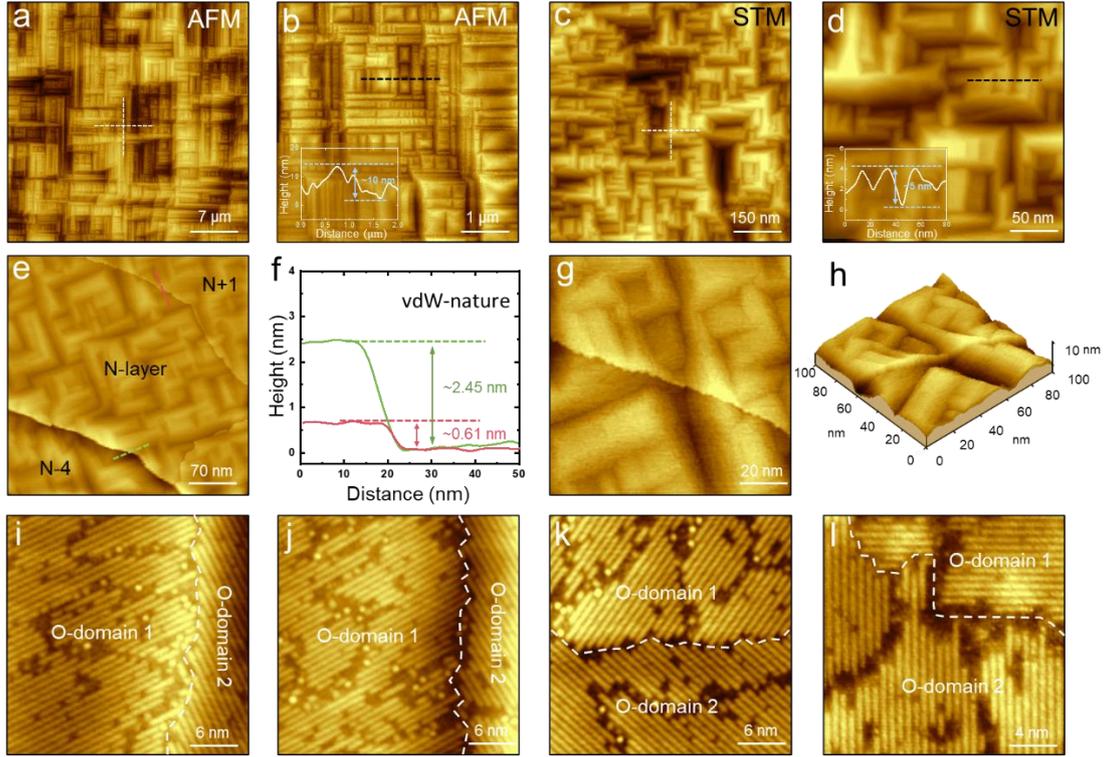

**Figure 2. Microscale AFM and nanoscale STM measurements of FePd$_2$Te$_2$.** (a) Micrometer-scale AFM topography of the freshly cleaved crystal with orthogonal characteristics. (b) Typical AFM topography of the mesoscale orthogonal corrugated regions. Inset: Line profile along the black dashed line in (b) with vertical heights of ~10 nm. (c) Nanometer-scale STM image with orthogonal characteristics. (d) Typical STM image of the nanoscale orthogonal corrugated domains. Inset: Line profile along the dashed line in (d) with vertical heights of ~5 nm. (e,f) Typical STM image (e) and height line profiles (f) for the surface step edges, demonstrating the 2D-layered nature of the corrugated crystal. (g,h) 2D- (g) and 3D-view (h) of STM image with a single-layer step edge. (i-l) STM images of O-domains with the ridge (i), valley (j,k) and flat (l) boundaries, which are highlighted by the white dashed lines. Scanning parameters: (c,d) V=-0.2 V, I=-100 pA; (e,g) V=-0.6 V, I=-200 pA; (i-l) V=-0.3 V, I=-200 pA.

Figure 2a and 2b present the large-scale and zoom-in AFM topography of the fresh-cleaved FePd$_2$Te$_2$ surface, in which the complex orthogonal (marked by the dashed cross) and corrugated (shown by the line profile) morphology are directly resolved at the micrometer-scale and consistent with the optical characterizations. Figure 2c and 2d present the typical large-scale and high-resolution STM topography images, in which the similar orthogonal and corrugated features are also observed at the nanometer-scale. It is clear that the above observed orthogonal features are self-similar hierarchical across nanoscale-to-microscale, additional

AFM and STM images are presented in Figure S5, which is never observed in 2D layered material but occasionally occurs in specific 3D bulk materials according to our knowledge.

Even with the prominent surface corrugations, the layered nature of $FePd_2Te_2$ as atypical 2D materials still can be clearly identified by the distinct single- and four-layer step edges, as shown in Fig. 2e and 2f. It has been confirmed the van der Waals-like stratification of the crystal. The determined thickness of single-layer is ~0.6nm, consistent with the XRD-determined structural model. The high-resolution 2D- and 3D-view STM image with the step edge, as shown in of Fig. 2g and 2h, directly indicate the continuity of surface corrugations across the step edges. It is clearly that the corrugations of each layer are consistent and transfer through their strong interlayer bonding. More STM images (Fig. S6) further confirm the consistency of the corrugations of each layer. This observation is critical, as it demonstrates that the complex orthogonal and corrugated morphology is not an artifact confined to a single layer single terrace or a result of local reconstruction.

Figure 2i and 2k show the atomic-scale STM images for the typical ridges and valleys areas of the mountain-like corrugations, in which the chain-like structures are clearly resolved. The marked ridge- and valley-domain walls separate the two stitched O-domains with orthogonal chains. The FFT and Fourier-filtered images of O-domain (Fig. S7) further confirm the orthogonality of the Fe-chains. It consistent with our proposed model for the structural relaxations due to the anisotropic mechanical properties (Fig. S4). Another two kinds of O-domain walls are also shown in Fig. 2k and 2l with tiny and negligible valley- and ridge-corrugations. Most O-domain boundaries are along the valley or ridges of surface corrugations, demonstrating the intertwined atomic structures, nanoscale domains and microscale morphology via the anisotropic Fe-chains.

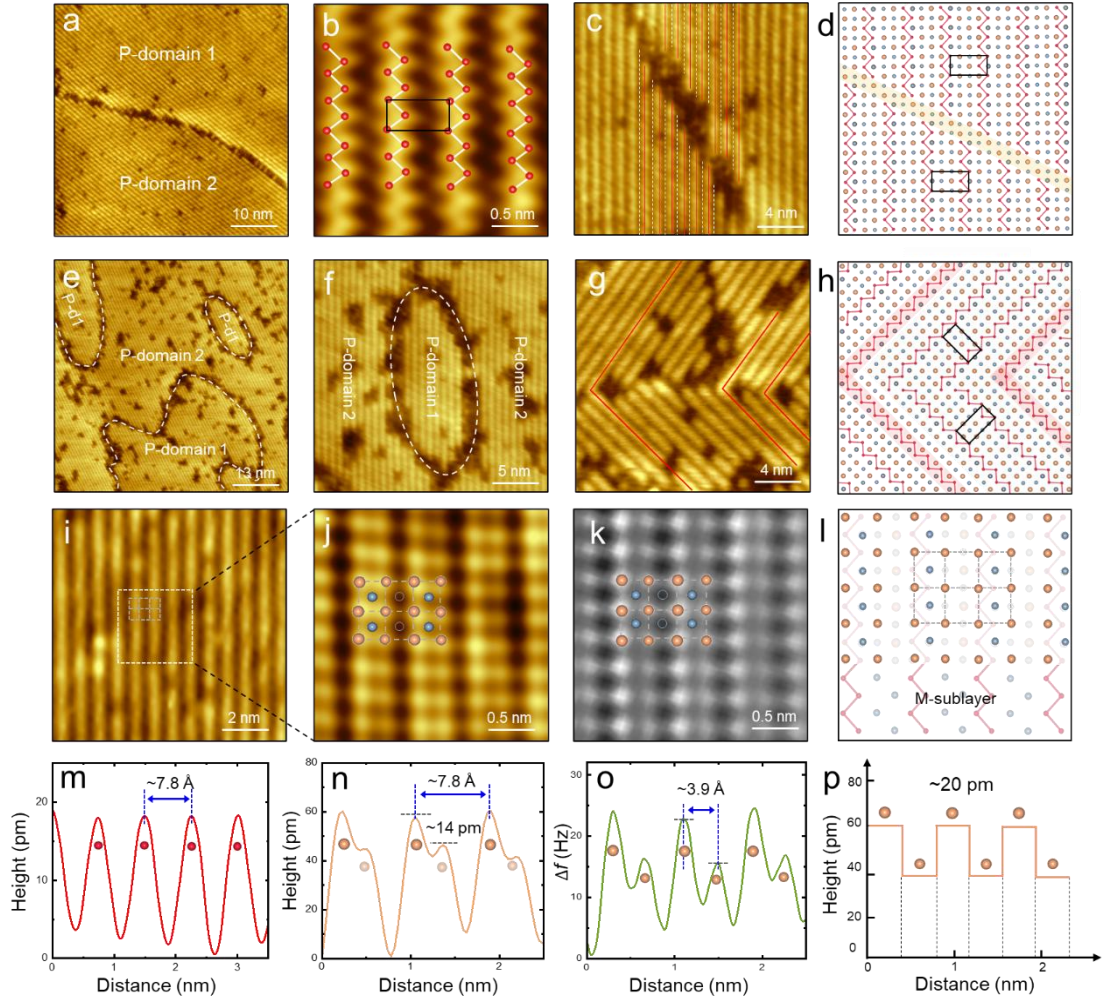

**Figure 3. Phase-domain and atomic-scale structures of the intact FePd$_2$Te$_2$ layer.** (a) STM image of the pristine surface with single O-domain made of two P-domains. (b) High-resolution STM image of Fe-zigzag chains with its overlaid model. (c,d) STM image of the two P-domains (c) and its schematic model (d) with the broken boundary. (e,f) Large-scale (e) and zoom-in (f) STM images of the P-domains with the highlighted boundaries. (g,h) High-resolution STM image of the O-domain boundary (g) and its schematic model (h). (i,m) High-resolution STM image (i) and line profile of the Fe-chains (m). (j,n) Atomic-resolution STM image (j) and line profile (n) of the top-Pd$_{1/2}$Te-sublayer, exhibiting two kinds of Te atoms with slight apparent height difference. (k,o) Atomic-resolution nc-AFM image (k) and line profile of the Te atoms (o) in the top-Pd$_{1/2}$Te-sublayer. (l) Schematic model of the top-Pd$_{1/2}$Te-sublayer (top-panel) and middle-FePd-sublayer (bottom-panel). (p) Schematic height profile of the Te atoms within the top-sublayer. Scanning parameters: (a,i) V=-0.3 V, I=-200 pA; (b,j) V=-0.7 V, I=-140 pA ; (c) V=-0.7 V, I=-200 pA ; (e-g) V=-0.5 V, I=-100 pA.

Based on its specifical crystal structure shown in Fig. 1, it is inferred that the CDW-like phase-domains (P-domain) will also exist and should be observed in the STM images of $FePd_2Te_2$. Different from the large corrugated O-domains, the P-domains only are found within the O-domains, as shown in Fig. 3a, in which two P-domains coexist within a single O-domain. The higher-resolution STM image of Fig. 3b directly shows the zigzag structure of Fe-chains within the middle-sublayer of $FePd_2Te_2$. A zoomed-in STM images of P-domain boundary is presented in Fig. 3c, in which the Fe-chains are marked with red solid lines and separated by the white dashed lines. This clearly shows that the Fe-chains of P-domains shift one Te atomic period across the P-domain boundary, as schematically shown in the structural model of Fig. 3d. It is noted that t the Fe-chains are completely disconnected at the P-domain boundary. This finding is also corroborated by the observed P-domain boundary marked with white dashed lines in both the larger-scale (Fig. 3e) and zoomed-in (Fig. 3f) P-domain STM images.

Meanwhile, unlike the corrugated O-domain in Fig. 2a, the P-domains exhibit no significant vertical height difference (Fig. S9). As shown in Fig. S10, the anisotropic Fe-chains contribute the different mechanical properties (Young's modulus) across the boundary of O-domain but not across the boundary of P-domains. Due to the absence of internal strain, no corrugated regions formed across the P-domain boundaries during the structural relaxation, indicating the difference between P-domains and O-domains. Furthermore, the distinction between the two domains also lies in the Fe-chains at their boundary. The STM image of the zoomed-in O-domain (Fig. 3g) clearly shows that some Fe chains can directly connect and rotate by 90° (highlighted by red solid lines) across the O-domain boundary. The corresponding structural model is presented in Fig. 3h.

The existence of these two kinds of domain structures relies entirely on the unique Fe-chain structures of $FePd_2Te_2$. Next, we further conducted atomic-resolution studies on the surface layer. Figures 3i and 3m display high-resolution STM images and line profiles of the Fe-chains, in which the Fe-chains appear as the parallel bright lines with uniform distribution and consistent height. Figure 3j and 3n shows its corresponding atomic-resolution STM images and line profiles of the top-$Pd_{1/2}$Te sublayer at the same area, in which two kinds of Te atoms are observed with slight apparent height difference. This consistent with the atomic structure diagram in Fig. 1a and directly shown with the overlaid model in Fig. 3j. Furthermore, the atomic-resolution nc-AFM image (Fig. 3k) and line profile of Te atoms (Fig. 3o) provide additional confirmation of the slight height difference between Te atoms. It should be noted that the STM image in Fig. 3j reflects a convolution of surface geometry and local electronic states. Therefore, the measured spacing between adjacent Te atoms may not correspond strictly to the real atomic distance. A schematic model of the top-$Pd_{1/2}$Te sublayer and middle-FePd sublayer is depicted in Fig. 3l. The Te atoms in the top-$Pd_{1/2}$Te-sublayer are marked with dashed squares, and the Pd-atoms and Pd-voids of the top-sublayer are also resolved and marked with the blue balls and empty circles. It is also noted that the slight height difference of Te atoms represents the underlying respective Fe-chains and Pd-chains within the middle-FePd-sublayer. Figure 3p displays the schematic height profile of the Te atoms within the top-sublayer, illustrating the unique characteristics of Te atoms in the $Pd_{1/2}$Te-sublayer.

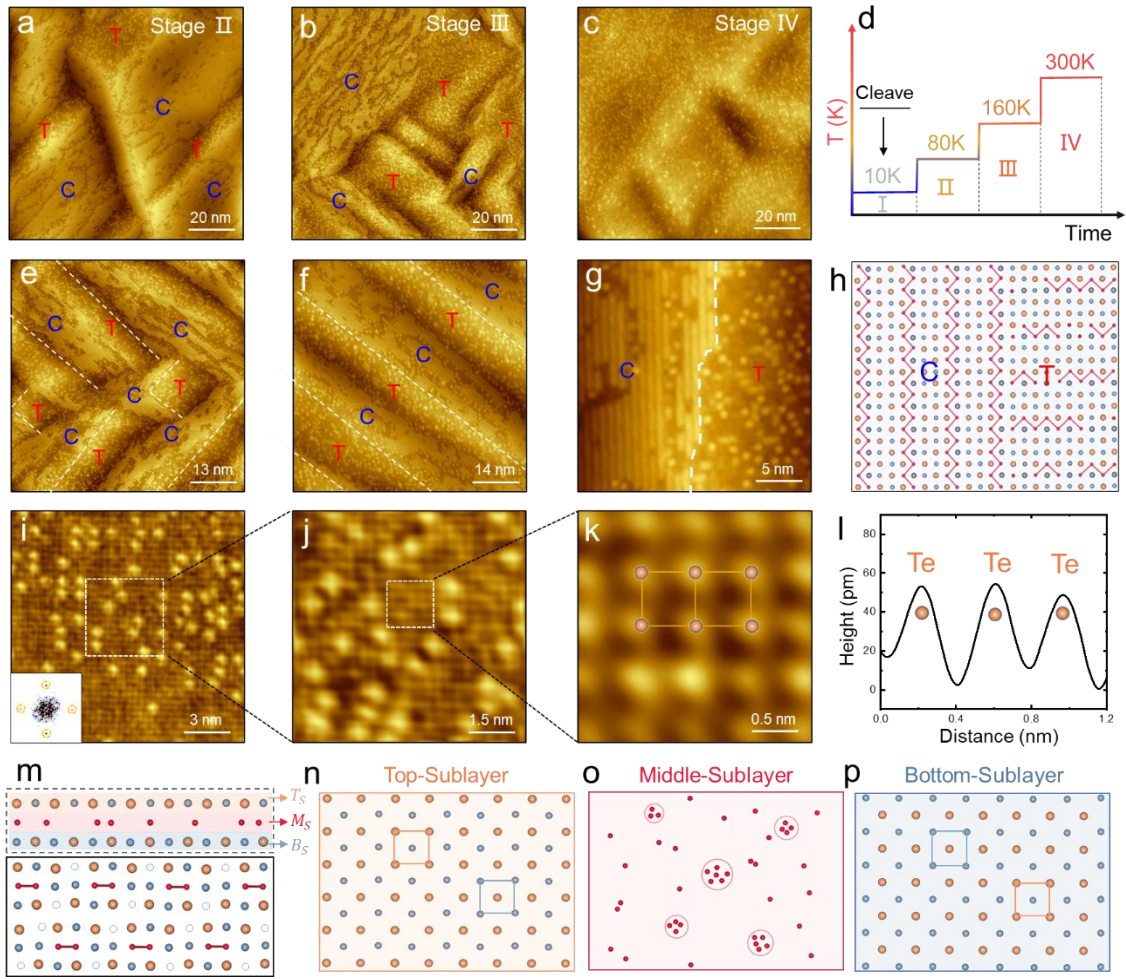

**Figure 4. Temperature-dependent structural transitions of the FePd$_2$Te$_2$ surface layer.** (a-d) STM images of the surface layer at the stage-II (a, ~80K), stage-III (b, ~160K) and stage-IV (c, ~300K) from the initial intact stage-I (Figure 3a, ~10K), sequentially obtained during the schematic procedure (d). (e,f) STM images (stage-II) of the C and T regions with higher and lower stabilities, indicated by their relatively intact and broken Fe-chains respectively. (g,h) High-resolution STM image and schematic model for the intact and broken Fe-chains at the C and T regions. (i,j) Large-scale (i) and high-resolution (j) STM images of the surface layer after thermal-induced structural transition, in which the relative uniform Te atoms and bright spots (possible Fe atoms/clusters) are resolved. (k,l) Atomic-resolution STM image (k) and line profile (l) of the top-sublayer Te atoms, different from Fig. 3j. (m-p) Proposed schematic models of the surface layer (m) after structural transition and its top-sublayer (n, PdTe), middle-sublayer (o, Fe atoms/clusters) and bottom-sublayer (p, PdTe). Scanning parameters: (a-c) V=-0.5 V, I=-100 pA; (e-g) V=-0.2 V, I=-200 pA; (i-k) V=-0.1 V, I=-200 pA.

Due to the anisotropic Young's modulus, the formation of compressive (C) and tensile (T) regions can occur under the inherent internal strains, which is related to the corrugated O-domains. After the low-temperature cleavage, the $FePd_2Te_2$ samples can sequentially undergoes four different distinct stages (Fig. 4d). At the Stage-I (~10K, directly after low-temperature cleaving), as shown in the STM images of Fig.2 and Fig.3, all the O-domains show the intact Fe-chains. With the increased sample temperature at ~ 80K (Stage-I), only partial corrugated regions (marked by red **T**) show the decomposed Fe-chains as the distributed bright spots in the STM images (Fig. 4a). At the Stage-II (~160K), the Fe-chains in the remained regions (marked by blue **C**) also gradually decomposed (Fig. 4b). At the Stage-IV (~300K), all of the Fe-chains completely decomposed into the distributed bright spots. The specific thermal-stability of different regions should be related to their internal strain conditions. Figure 4e and 4f display the typical STM images at the stage-II, the interwoven C and T domains clearly display the intact and decomposed Fe-chains, respectively, indicating their higher and lower thermal stabilities. Figure 4g and 4f shows the high-resolution STM image and schematic model for the intact and decomposed Fe-chains at the C and T regions. This temperature-dependent structural transition occurs exclusively at the surface layer, and the underlying layers retains their intact structures (Fig. S12).

The temperature-dependent structural transition at the surface layers is temperature-irreversible, and their transited structures with the decomposed Fe-chains can be further atomically investigated at ~ 10K. Figure 4i and 4j show large-scale and high-resolution STM images of the surface layer after thermal-induced structural transition. It is clear that no pristine parallel chains but the randomly distributed bright spots (possible Fe atoms/clusters) are observed with the uniform Te atoms in the square configuration. The zoomed-in STM images reveal the atomic-resolution image (Fig. 4k) and line profile (Fig. 4l) of the top-sublayer Te atoms. Different from Fig. 3j, the top-sublayer Te atoms exhibit relative uniform height after the structural transition. The proposed structural model of surface layer after the transition are schematically displayed in Fig. 4m. The detailed structural transition processes are schematically shown in Fig. S13. After the cleavage, the decreased interlayer interactions of surface layer weaken its intralayer structural stabilities. With the thermal energies, the Pd-atoms of middle-sublayer can be shifted to fill the Pd-voids at the top- and bottom-sublayers to form the stable top- and bottom-PdTe-sublayers (Fig. 4n and Fig. 4p). The remained Fe-chains within the middle-sublayer are structurally unstable without the confinement effect of Pd-chains and gradually decomposed into disordered Pd atoms/clusters (Fig. 4o) confined between the well-structured top- and bottom-PdTe-sublayers. By comparing the structural models before and after the transition (Fig. S14), we can conclude that the unique shifting of Pd-atoms/voids among the sandwich-like sublayers play the critical roles on the formation of intra-layer anisotropic Fe-chains and the inter-layer structural alignment via the strong interlayer bonding.

**Discussion**

This work demonstrates that the unique structure of the layered ferromagnet $FePd_2Te_2$ is fundamentally determined by the arrangement of Pd atoms and voids. At the atomic scale, these Pd atoms/voids exert a critical influence on the two kinds of Te atoms in the top-sublayer and the anisotropic 1D Fe-chains in the middle-sublayer, thereby controlling interlayer interactions. This atomic order dictates the formation of nanoscale self-organized domains, driven by a chain-orientation-dependent mechanical coupling effect. Ultimately, these domains assemble into orthorhombic corrugated morphologies at the mesoscale. Through combined scanning tunneling microscopy and atomic force microscopy observations, we directly visualized this cross-scale correlation, revealing how local atomic arrangements dictate structural order at both nanoscale and micrometer scales. Furthermore, we observe an intriguing strain- and temperature-dependent reconstruction of the surface layer. The atoms in the surface sandwich-layer keep long-range ordered four-fold symmetry in the top- and bottom-sublayers, while in contrast, the Fe-atoms in the middle-sublayer form a completely disordered cluster-like structure. The discovery of this unique order-disorder stacking structure and its formation process provide insights and new strategies in engineering 2D materials in atomic-scale.

This unique order-disorder stacking structure may explain the structural phase transition in $FePd_2Te_2$ bulk crystals that discovered recently [34]. Penacchio et al. report the bulk $FePd_2Te_2$ have a structure phase transition near 420 K and might form a new crystal structure with four-fold symmetry, but the detailed crystal structure is unresolved. Since the order-disorder stacking surface structure discovered in this work also has four-fold symmetry and tend to form at higher temperatures, so an interesting speculation is that this surface structural modulation may extend to the whole bulk near 420 K. If so, new opportunities in modulating the atoms in the bulk near room temperature will be provided. Ruiz et al. reported that substitution atoms of Co and Ni can effectively modulate the exchange coupling and anisotropy properties of $FePd_2Te_2$ through DFT calculations[42]. Mechanical strain serves as an effective means to regulate the magnetic properties of $FePd_2Te_2$ and $CoPd_2Te_2$, providing theoretical support for designing spintronic devices with customized magnetism through chemical and structural manipulation. This work further reveals that the distribution and occupancy of Pd atoms/voids are key factors determining the 1D Fe-chains structure and orthorhombic domains morphology. Based on this, we propose that methods such as atomic intercalation to alter van der Waals interlayer spacing, quenching treatment, element substitution, and designing specific element ratios can effectively regulate Pd atoms/voids arrangement and content, thereby directly influencing Fe-chains configuration and domains. This potentially provides a clear physical picture and a feasible material design platform for achieving fine-tuned control of magnetic anisotropy experimentally.

**Conclusions**

Combined STM/AFM data direct confirms that the unique hierarchical structure of FePd$_2$Te$_2$ originates from the atomic-scale configuration of Pd atoms/voids. This arrangement simultaneously regulates the two kinds of Te atoms and anisotropic Fe-chains, thereby driving the self-assembly of nanoscale domains and ultimately forming an orthorhombic corrugated morphology at the mesoscale. Furthermore, we observe a unique coexistence of ordered and disordered phases in the surface layers: top-/bottom-sublayers consist of ordered PdTe sublayers, intercalated with disordered Fe-sublayers. This distinctive stacking structure arises from its evolution with strain and temperature. Our work establishes a novel cross-scale engineering approach for controlling the structure and magnetic anisotropy of two-dimensional materials.

## Materials and Methods

**Sample preparation and characterization.**

Single crystals of $FePd_2Te_2$ were grown by melting stoichiometric elements. Iron, palladium, and tellurium powder were mixed and ground in a molar ratio of 1:2:2. Deviation from this ratio would lead to the unsuccessful growth of large single crystals or the target phase. Then the mixtures were placed in an alumina crucible and sealed in a quartz tube under vacuum conditions. The entire tube was heated in a box furnace to 800 °C and held at that temperature for 2 days. Then it was cooled to 600 °C at a rate of 2 °C/h followed by annealing at this temperature for 2 days before being furnace-cooled to room temperature. In addition, we found that quenching the samples at 600 °C would improve the crystal quality, as revealed from X-ray characterization. However, direct quenching in the initial growth process seems to break the large crystal into small pieces. By reannealing and quenching the large single crystal grown by the initial furnace-cooled method, one could obtain crystals with both large size and good quality.

**Structure and Composition Characterization.**

The structure of $FePd_2Te_2$ crystals was analyzed at 273 K by XRD (D8 ADVANCE, Bruker) equipped with multilayer mirror monochromatized Mo Kα (λ = 0.71073 Å) radiation. The elemental composition and distribution were evaluated by EDS (X-MaxN 50 $mm^2$, Oxford Instruments) in the SEM.

**AFM measurements**

The AFM experiments were performed using a commercial atomic force microscope (Park, NX10) equipped with a commercial topography tip (Nanosensors, AC160TS, Quality factor ~500 at room temperature). The scanning probe system was operated at the resonance frequency of the topography tip, ~301 kHz. The AFM images were acquired in non-contact mode.

**STM and nc-AFM measurements.**

High-quality $FePd_2Te_2$ crystals were cleaved at room temperature in ultrahigh vacuum at a base pressure of $2\times10^{-10}$ Torr, and directly transferred to the cryogen-free variable-temperature STM system (PanScan Freedom, RHK). Chemically etched Pt-Ir tips were used for STM measurement in constant current mode. The tips were calibrated on a clean Ag(111) surface. Gwyddion was used for STM data analysis. $FePd_2Te_2$ crystals were cleaved at liquid nitrogen temperature in ultrahigh vacuum .The nc-AFM measurements were performed in a commercial LT-STM system (LT-STM/AFM, CreaTec) equipped with an STM/qPlus sensor at 4.5K. The nc-AFM images were recorded by measuring the frequency shift of the qPlus resonator (sensor frequency $f_0$=30kHz, Q= 53000) in constant-height mode with an oscillation amplitude of 200 pm.

**Acknowledgments**

This project was supported by the National Key R&D Program of China (MOST) (Grant No. 2023YFA1406500. 2024YFA1207700), the National Natural Science Foundation of China (NSFC) (No. 92477128, 92580137,92477205, 12374200, 11604063, 11974422, 12104504, 12504203, 62488201), the Fundamental Research Funds for the Central Universities and the Research Funds of Renmin University of China (No. 21XNLG27), and High-level Talent Research Start-up Project Funding of Henan Academy of Sciences (Project 20251827003). Y.Y. Geng was supported by the Outstanding Innovative Talents Cultivation Funded Programs 2023 of Renmin University of China. This paper is an outcome of "Two-dimensional anisotropic series of materials $FePd_{2+x}Te_2$: a structural modulation study from the atomic scale to the mesoscopic scale " (RUC25QSDL128), funded by the Qiushi Academic-Dongliang Talent Cultivation Project at Renmin University of China in 2025.

**Author contributions**

L.L., C.P. and Z.C. conceived the research project. M.W. performed the STM experiments and analysis of STM data. , S.M. and X.P performed the AFM measurements. B.X. grew the single crystals. Y.G., S.M., F.P., R.X., L.H. and W.J., H.G. helped in the experiments. M.W. and Z.C. wrote the manuscript with inputs from all authors.

**Competing Interests**

The authors declare no competing financial interests.

**Data Availability**

The authors declare that the data supporting the findings of this study are available within the article and its Supplementary Information.